\documentclass[aps,prl,twocolumn,showpacs,superscriptaddress,groupedaddress]{revtex4}  
\usepackage{subfigure}
\usepackage{graphicx}  
\usepackage{dcolumn}   
\usepackage{bm}        
\usepackage{amssymb}   
\usepackage{color}
\usepackage{slashed}
\usepackage{soul}
\usepackage{times}
\usepackage{xspace}

\def\lres{L_{\rm \small  {res}}}

\def\jraa{R_{\tiny{\rm AA}}^{\rm jet}}
\def\hraa{R_{\tiny{\rm AA}}^{\rm had}}

\def\aSC{{\kappa_{\rm sc}}}

\newcommand{\pythia}{{\sc Pythia}\xspace}

\begin{document}


\hspace{5.8in} \mbox{MIT-CTP-5050}

\title{A Simultaneous Description of Hadron  and Jet Suppression in Heavy Ion Collisions}
\date{\today}

\author{J. Casalderrey-Solana}
\affiliation{Departament de F\'\i sica Qu\`antica i Astrof\'\i sica \& Institut de Ci\`encies del Cosmos (ICC), Universitat de Barcelona, Mart\'{\i}  i Franqu\`es 1, 08028 Barcelona, Spain}
\affiliation{Rudolf Peierls Centre for Theoretical Physics, University of Oxford, Clarendon Laboratory,
Parks Road
Oxford OX1 3PU United Kingdom}
\author{Z. Hulcher}
\affiliation{Cavendish Laboratory, University of Cambridge, JJ Thomson Ave, Cambridge CB3 0HE, UK}
\author{G. Milhano}
\affiliation{LIP, Av. Prof. Gama Pinto, 2, P-1649-003 Lisboa , Portugal}
\affiliation{Instituto Superior T\'ecnico (IST), Universidade de Lisboa, Av. Rovisco Pais 1, 1049-001, Lisbon, Portugal}
\author{D. Pablos}
\affiliation{Department of Physics, McGill University, 3600 University Street, Montr\'eal, QC, H3A 2T8, Canada}
\author{K. Rajagopal}
\affiliation{Center for Theoretical Physics, Massachusetts Institute of Technology, Cambridge, MA 02139 USA}

\begin{abstract}
We present a global fit to all data on the suppression of high energy jets and high energy hadrons in the most central heavy ion collisions at the LHC for two different collision energies, within a hybrid strong/weak coupling quenching model. Even though the measured suppression factors for hadrons and jets differ significantly from one another and appear to asymptote to different values in the high energy limit, we obtain a simultaneous description of all these data after constraining the value of a single model parameter. We use our model to investigate the origin of the difference between the observed suppression of jets and hadrons and relate it, quantitatively, 
to the observed modification of the jet fragmentation function in jets that have been modified by
passage through the medium produced in heavy ion collisions. 
In particular, the observed increase in the fraction of hard fragments in medium-modified jets, which indicates that jets with the fewest hardest fragments lose the least energy, corresponds quantitatively to the observed difference between the suppression of hadrons and jets.
We argue that a harder fragmentation pattern for jets with a given energy after quenching 
 is a generic feature of any mechanism for the interaction between jets and the medium that they traverse  that yields a larger suppression for wider jets. We also compare the results of our global fit to LHC data to 
measurements of the suppression of high energy hadrons in RHIC collisions, and find that with its parameter chosen to fit the LHC data
our model 
is inconsistent with the RHIC data 
at the $3\sigma$ level,
suggesting that hard probes interact more strongly with the less hot quark-gluon plasma produced at RHIC.
\end{abstract}


\maketitle


\textit{Introduction.}  
One of the most striking observations of the heavy ion physics programs of both RHIC and the LHC is the suppression in the measured
yield of high-energy jets and hadrons
in ultrelativistic nucleus-nucleus collisions relative 
to the expected yield if these collisions were just an incoherent superposition of independent nucleon-nucleon collisions.
This phenomenon, which is generically known as jet quenching, is a direct consequence of the 
energy loss experienced by the high-energy partons that form jets and subsequently decay into hadrons
%
as these partons traverse the strongly coupled quark-gluon plasma (QGP) produced in the same heavy ion collisions.
Since such parton-medium interactions have the potential to provide tomographic information about the microscopic properties of QGP, the study of the suppression patterns of different 
energetic probes has been the subject of considerable experimental and theoretical research. For recent reviews, see Refs.~\cite{Mehtar-Tani:2013pia,Qin:2015srf,Connors:2017ptx,Busza:2018rrf}.

RHIC pioneered the measurement of the suppression in the measured yield of high energy hadrons,
data that we shall return to later.
The LHC heavy ion physics program has provided us with a wealth of data that characterizes the 
measured yields of both high-energy hadrons and high-energy jets, with larger and larger data sets anticipated as the LHC luminosity is increased in the early 2020s~\cite{Jowett,CMSfuture}.  The calorimetric measurement of high energy jets at RHIC is a central goal of the sPHENIX detector~\cite{Adare:2015kwa}, with high statistics measurements anticipated in the early 2020s also.
One of the important questions posed by today's data is how to understand 
the basic empirical feature
that high energy hadrons are less suppressed than jets with the same or higher energies.
In this paper, we shall use an analysis couched within a specific model for jet quenching to elucidate a generic explanation for this characteristic feature of the observed data. In so doing, we shall find evidence in support of generic aspects of the parton-medium interaction responsible for jet quenching that are necessary to explain the observed systematics.

Jets are the sprays of hadrons produced as an energetic parton from an elementary parton-parton collision 
 fragments into a shower of partons which ultimately color-neutralize into hadrons.
The partons
that we are discussing have energies that are orders of magnitude
larger than the typical thermal scales that characterize the QGP produced  in the same collisions.
Nevertheless, as these energetic jets traverse the QGP they lose energy, and their properties 
 are modified. Because the production rate for jets in elementary collisions drops rapidly with increasing jet energy, which is to say because 
   the jet production spectrum is steeply falling, jet energy loss implies a suppression in the yield of jets with a given energy relative to what it would have been in the absence of any medium.
   Since high-energy hadrons originate from partons within jets, this suppression in turn translates into the suppression of high energy hadrons.
 %
While
the suppression pattern of jets and hadrons are therefore related to each other,
even in LHC heavy ion collisions with the highest center of mass energy per nucleon pair achieved to date, $\sqrt{s_{\rm NN}}=5.02$~TeV, the production of jets with a given energy is more suppressed than the production of hadrons with the same energy and, it seems, the production of the highest energy jets that have been measured is more suppressed than the production of the highest energy hadrons that have been measured~\cite{Khachatryan:2016odn,ATLAS:2017rmz,Aaboud:2018twu}. 
We shall provide an explanation of these basic systematic features in the data.

The essence of our explanation originates from
the fact that when one selects a hadron with a specified high $p_T$ (for example 
many tens of GeV), although this hadron originates from a jet
the population of jets that is selected in this way is not typical, 
not representative of the 
generic population of jets selected calorimetrically by finding sprays of particles
whose total $p_T$ has a specified value.  In particular, the population of jets that dominate
the production of high-$p_T$ hadrons will have a nongeneric probability distribution for their angular widths, as we now explain.
The fact that the jet spectrum is steeply falling as a function of $p_T^{\rm jet}$ means that
a hadron with any given large $p_T^{\rm hadron}$ is most likely to be a hadron that
carries a large fraction of all of the energy of the jet in which it finds itself.
 This is because if the hadron carried a smaller fraction of the total energy of its jet, that would mean its jet had a larger total $p_T^{\rm jet}$, which is less likely because the spectrum is steeply falling. 
Consequently, high-energy hadrons belong 
to an unrepresentative subset of jets which happen to fragment such that they contain very few hard partons outside the jet core, and are consequently narrow in their angular extent. 
If jets of this type lose less energy than typical jets, this explains why
the yield of high-$p_T$ hadrons is less suppressed than the yield of high-$p_T$ jets.
%
%
%
Furthermore, 
if wider jets that contain more partons at large angles lose more energy 
due to quenching, this together with the steeply falling
jet spectrum means that quenching makes it more likely to find narrow jets with fewer 
harder hadrons, since they are the ones that lost less energy.
So, the same physics that explains why hadrons are more
suppressed than jets also yields a modification to the jet fragmentation function.



There are many extant models for the jet-medium interaction in which 
hard fragmenting, narrow, jets lose less energy than typical, wider, jets. Examples include models
based entirely on perturbative QCD~\cite{Milhano:2015mng}, strong coupling models built entirely using holographic techniques~\cite{Rajagopal:2016uip,Brewer:2017fqy}, and the hybrid model that we shall employ~\cite{Casalderrey-Solana:2014bpa,Casalderrey-Solana:2015vaa,Casalderrey-Solana:2016jvj,Hulcher:2017cpt}.
Within the hybrid model, we shall show  that the same physics which
enhances the probability for finding hard fragments in jets that have traversed a droplet of
plasma explains the observed difference between the suppression of hadrons and jets.
Our  arguments imply that this should be the case, at least at a qualitative level in any model with the 
underlying feature that narrow jets lose less energy than wide jets.



\textit{The hybrid strong/weak coupling model.}
The physics of hard parton production and subsequent showering can be analyzed
perturbatively, with weakly coupled QCD, making it natural to develop weakly coupled analyses of parton energy loss in medium~\cite{Baier:1996kr,Zakharov:1996fv,Baier:1998kq,Gyulassy:2000er,Wiedemann:2000za,Wang:2001ifa,Arnold:2002ja,Salgado:2003gb,Jeon:2003gi,Jacobs:2004qv,Lokhtin:2005px,CasalderreySolana:2007zz,Zapp:2008af,Zapp:2008gi,Lokhtin:2008xi,Armesto:2009fj,Schenke:2009gb,Majumder:2010qh,CasalderreySolana:2010eh,DEramo:2012uzl,Mehtar-Tani:2013pia,Wang:2013cia,Zapp:2013vla,Ghiglieri:2015zma,He:2015pra,Blaizot:2015lma,Qin:2015srf,Chien:2015hda,Cao:2017zih,Arleo:2017ntr,DEramo:2018eoy}. 
A weakly coupled calculation is, however, not the natural starting point with which to describe
the hydrodynamic expansion of the droplets of strongly coupled QGP produced in RHIC and LHC
collisions, and may not be the best starting point for describing the physics of typical soft momentum exchanges between partons in a jet shower and this medium.
Complementarily, numerous qualitative
insights into the 
properties of the strongly coupled QGP can be obtained from gauge/gravity duality, which
has emerged as a tool with which to analyze the dynamics of 
droplets of strongly coupled gauge theory plasmas
that are similar to QGP.  (See Ref.~\cite{CasalderreySolana:2011us} for a comprehensive review.)
These 
methods cannot be employed to describe hard processes, such as jet production, which
are sensitive to short-distance physics where QCD is, in fact, weakly coupled. 
The wide  range of physical scales relevant for the interaction of hard probes  with the QGP 
medium make the understanding of jet and hadron suppression an interesting theoretical challenge.

To address this challenge, a phenomenological hybrid model that exploits the separation between jet and medium scales and brings together the relevant description of the physics at each energy scale was developed in Refs.~\cite{Casalderrey-Solana:2014bpa,Casalderrey-Solana:2015vaa,Casalderrey-Solana:2016jvj,Hulcher:2017cpt}. 
In the model, partons generated in a hard collision, whose production is well described by diagrammatic techniques in perturbative QCD, relaxes its virtuality $Q$ down to the 
hadronization scale $\Lambda_{\rm QCD}$ through successive splittings, following DGLAP evolution equations as implemented in \pythia. 
Given the scale decoupling throughout most of the showering process between the virtuality and the medium temperature $T$, $Q \gg T$, 
we leave the splittings within the shower unmodified.
Simultaneously to the splitting processes, the partons in the developing shower traverse an expanding cooling droplet of QGP and we must treat their interactions 
with the medium in which they find themselves nonperturbatively, since $T \sim \Lambda_{\rm QCD}$. This 
motivates modelling the energy degradation of the partons within the jet shower 
by applying  geometric intuitions from holography. 
Via gauge/gravity duality, the energy loss of a quark traveling through the   hot plasma of  $\mathcal{N}=4$ supersymmetric Yang-Mills (SYM) theory at infinitely strong coupling and large-$N_c$ is given by~\cite{Chesler:2014jva,Chesler:2015nqz}
\begin{equation}\label{CR_rate}
\left. \frac{d E_{}}{dx}\right|_{\rm strongly~coupled}= - \frac{4}{\pi} E_{\rm in} \frac{x^2}{x_{\rm therm}^2} \frac{1}{\sqrt{x_{\rm therm}^2-x^2}} \quad , 
\end{equation}
where $E_{\rm in}$ is the parton's initial energy, and $x_{\rm \small therm}=(E_{\rm in}^{1/3}/T^{4/3})/2 \aSC$ is the maximum distance that a parton with this energy can travel through the strongly coupled plasma. The ``lost'' energy and momentum  becomes part of the droplet of  QGP, generating a wake therein that will follow a hydrodynamic evolution, 
eventually decaying into soft hadrons 
that, by momentum conservation, carry the ``lost'' momentum in the direction of the original jet~\cite{Casalderrey-Solana:2016jvj}. While in $\mathcal{N}=4$ SYM the value of $\aSC\approx 1$ has been determined, its value must be less in the QGP of QCD which has fewer degrees of freedom.
One of the assumptions of the hybrid model~\cite{Casalderrey-Solana:2014bpa,Casalderrey-Solana:2015vaa} 
is that all the differences between the interactions of jets with this strongly coupled
plasma and the QGP of QCD can be accounted for by varying this parameter. In this way, $\aSC$ becomes the principal free parameter in the model, controlling the degree of parton energy loss. We
shall determine its value by fitting to data.

\begin{figure}
\hfill
\includegraphics[scale=1.2]{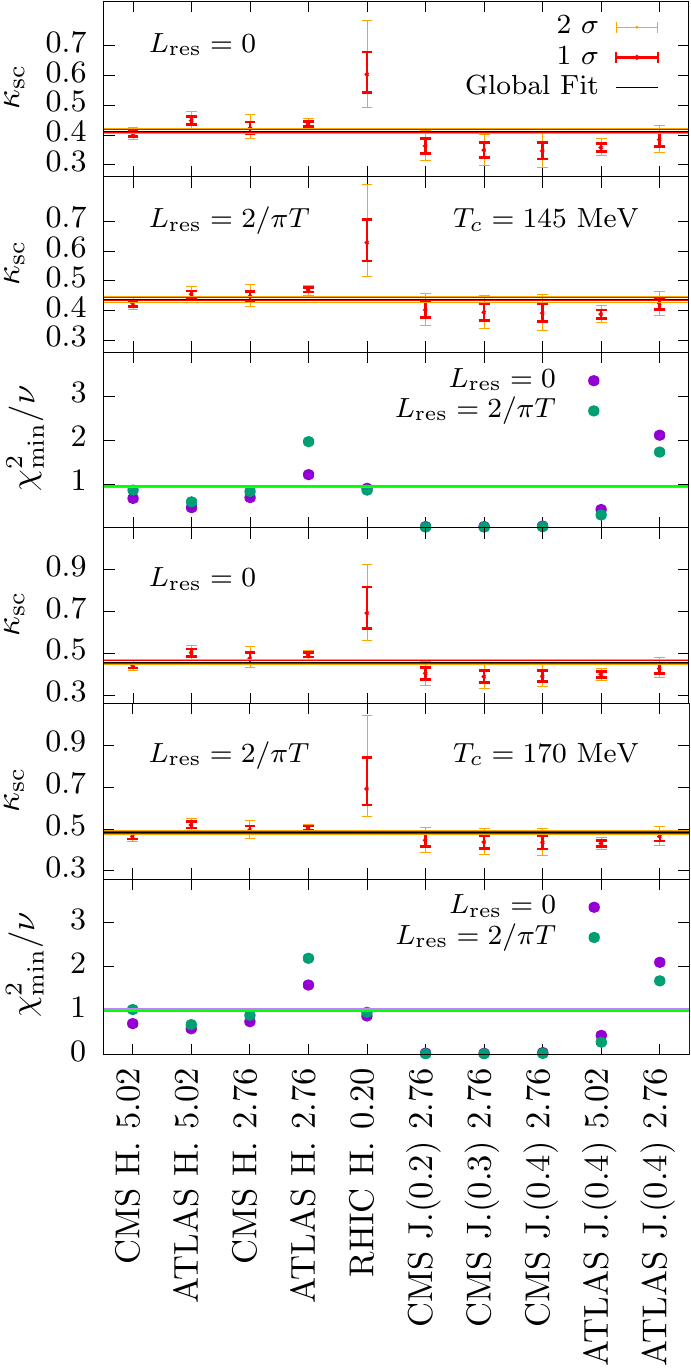}
\caption{\label{fig:fits} Results for the best values of $\aSC$ for $T_c=145$ MeV (first three panels) and $T_c=170$ MeV (last three panels). 
The individual red (orange) error bars show the $1\sigma$ ($2 \sigma$) uncertainties in the
value of $\aSC$ obtained by fitting separately to each one of ten data sets, nine from the LHC and one from RHIC.  ``H'' stands for charged hadrons (LHC, PbPb collisions, $\sqrt{s_{\rm NN}}$ specified in TeV) or $\pi^0$ (PHENIX, AuAu collisions, $\sqrt{s_{\rm NN}}$ again specified in TeV) in the 0-5\% centrality bin, while ``J'' stands for calorimetrically reconstructed jets, with the anti-$k_t$ radius~\cite{Cacciari:2008gp} in parentheses, in the 0-10\% centrality bin.
First panel of each set of three corresponds to $\lres=0$, second one to $\lres=2/(\pi T)$, and the third panel shows the goodness ($\chi^2$ per degree of freedom) of each fit. The horizontal red (orange) lines show the $1\sigma$ ($2 \sigma$)   range of values of $\aSC$ obtained via a global fit to all nine LHC data sets, and the (almost indistinguishable) purple and green horizontal lines show the goodness of these global fits for $\lres=0$ and $\lres=2/(\pi T)$}
\end{figure}

Despite its simplicity, this hybrid strong/weak coupling model has been very successful at describing inclusive jet and di-jet observables \cite{Casalderrey-Solana:2014bpa}, photon-jet and $Z^0$-jet observables \cite{Casalderrey-Solana:2015vaa}, as well as the more differential jet sub- and super-structure observables \cite{Casalderrey-Solana:2016jvj}. These validations of the core features of the model have turned it into a useful phenomenological tool with which to interpret the data, a tool 
which can be improved by adding additional physical phenomena and which therefore
allows for a detailed characterization of the impact that different physical phenomena can have on measurable observables.
One such extension is the analysis of finite resolution (or coherence)  effects in  Ref.~\cite{Hulcher:2017cpt}, which incorporates the fact that plasma cannot resolve  the internal structure of a parton shower with arbitrary precision, but can only interact independently with distinct excitations if they are separated by more than the plasma resolution length $\lres$. While this is a well studied phenomenon occurring both at weak \cite{MehtarTani:2011tz,CasalderreySolana:2011rz,CasalderreySolana:2012ef,Casalderrey-Solana:2015bww} and strong coupling \cite{Casalderrey-Solana:2015tas}, the analysis in Ref.~\cite{Hulcher:2017cpt} is the first exploratory study of these effects within a jet quenching Monte Carlo. In this letter we will also explore the sensitivity of suppression data to coherence phenomena, doing our global fit for $\lres=0$ and for the reasonable (see Ref.~\cite{Hulcher:2017cpt}) value $\lres=2/(\pi T)$.
As in all of Refs.~\cite{Casalderrey-Solana:2014bpa,Casalderrey-Solana:2015vaa,Casalderrey-Solana:2016jvj,Hulcher:2017cpt}, we shall do all our calculations for two different values of
 the temperature $T_c$ below which we turn off parton energy loss, $T_c=145$~MeV and 170 MeV, using this as a crude proxy by which to assess sensitivity to some systematic effects not included in the model.

\textit{A fit to hadron and jet suppression data.} 
We fix the free parameter of the model, $\aSC$, %
 by fitting to hadron and jet experimental data from  LHC (PbPb collisions at $\sqrt{s_{\rm NN}}=2.76$ \cite{CMS:2012aa,Khachatryan:2016jfl,Aad:2015wga,Aad:2014bxa} and $5.02$ TeV \cite{Khachatryan:2016odn,ATLAS:2017rmz,Aaboud:2018twu}) in the most central bins. The simulations rely on the event generator \pythia 8.230 \cite{Sjostrand:2014zea} for 
the production and DGLAP evolution of the shower and we include leading order nuclear parton distribution functions as parametrized in Ref.~\cite{Eskola:2009uj}.
%
 %
 The space-time picture of the shower is built 
 by assuming
 that the effective lifetime of each parton corresponds to $\tau=2E/Q^2$, with $E$ and $Q$ the energy and virtuality of that parton, respectively~\cite{CasalderreySolana:2011gx,Casalderrey-Solana:2014bpa}. When we choose $\lres=2/(\pi T)$ instead of $\lres=0$, this
has the effect of modifying these 
splitting times by delaying the time at which the QGP 
interacts with a pair of partons as distinct, separate, entities~\cite{Hulcher:2017cpt}.
The properties of the QGP needed to compute parton energy loss, namely the local temperature and fluid velocity, are read from hydrodynamic profiles for droplets of expanding cooling plasma that yield good 
descriptions of soft observables such as particle multiplicity and flow coefficients~\cite{Shen:2014vra}. These hydrodynamic simulations have a starting time of $\tau_0=0.6$ fm, before which we assume there is no energy loss. We stop applying energy loss when the local temperature goes below 
$T_c$, using two  different values for this quantity as noted above.
In order to estimate the contribution to the final hadron spectra coming from the wake generated by the passage of the jet through the plasma, as in  Ref.~\cite{Casalderrey-Solana:2016jvj}
we assume that the wake hydrodynamizes subject to momentum conservation, becomes a small perturbation to the bulk hydrodynamic flow, and yields a correction to the final hadron spectrum (obtained via the Cooper-Frye prescription~\cite{Cooper:1974mv}) that is also a small perturbation that can be linearized.
We perform the hadronization of the parton shower using the Lund string model present in \pythia, where, for simplicity, the color flow among the different partons is not modified.

We present in the six panels of Fig.~\ref{fig:fits} the results for the fits to the best values of $\aSC$ for the two different values of $T_c$ (first three panels for $T_c=145$ MeV, last three for $T_c=170$ MeV), and for $\lres=0$ and $2/(\pi T)$. 
The fits have been done in two different ways. First, the individual points with error bars are obtained by fitting the model, separately, to each of ten different sets of data using a standard $\chi^2$ 
analysis with different sources of experimental uncertainty (statistical, uncorrelated systematic, correlated systematic, and normalization) accounted for appropriately, as in Ref.~\cite{Adare:2008qa}.
And, second, the horizontal colored bands are obtained by performing a global fit to all nine LHC data sets. The uncertainty bands on these global fits correspond to the values of $\aSC$ for which $\chi^2=\chi_{\rm min}^2\pm1$ ($1 \sigma$) and $\chi^2=\chi_{\rm min}^2\pm4$ ($2 \sigma$).


We conclude from the global fit that our model can simultaneously describe data on the suppression of both hadrons and jets, yielding a satisfactory overall agreement between all sets of LHC data within the narrow range for $\aSC$ indicated by the global fit for either value of $\lres$ and $T_c$.
Although we certainly find no statistically significant preference for $\lres=0$ or $\lres=2/(\pi T)$ whatsoever, if we squint at Fig.~\ref{fig:fits} it appears that the agreement between the band of values of $\aSC$ found via the global fit and the jet suppression data looks slightly better for $\lres=2/(\pi T)$.
The global fit shows that this impression is not significant at present, but this impression --- and the goal of constraining the value of $\lres$ ---  motivates future higher statistics measurements of jet suppression.  
Note that although at fixed  $\aSC$ the effect of varying $\lres$ on jet suppression is significant, as noted in Ref.~\cite{Mehtar-Tani:2017web}, this dependence  
becomes rather weak after fitting the model parameter that controls the rate of parton energy loss --- in our case $\aSC$ which we determine via our  global fit. In any comparison between a perturbative analysis and data, fitting the value of the jet quenching parameter $\hat q$, as is appropriate and necessary, will have comparable consequences.


We see in Fig.~\ref{fig:fits} that the measurements of the suppression of $\pi^0$ yields in RHIC collisions~\cite{Adare:2008qa} 
favor a larger value of $\aSC$ than the one that we obtain from the global fit to LHC data, 
corresponding to a stronger coupling between energetic partons and the QGP that they traverse
in the lower temperature QGP produced at RHIC. This is in line with the finding of previous studies \cite{Horowitz:2011gd,Andres:2016iys}.
 However,
the distinction between the value of $\aSC$ preferred for RHIC and LHC collisions is 
not at the $5 \sigma$ level. 
This motivates future higher statistics measurements of both hadron and jet suppression at RHIC.
 It would also be interesting to extend this analysis to different centrality classes.

\begin{figure}
\hfill
\includegraphics[scale=0.72]{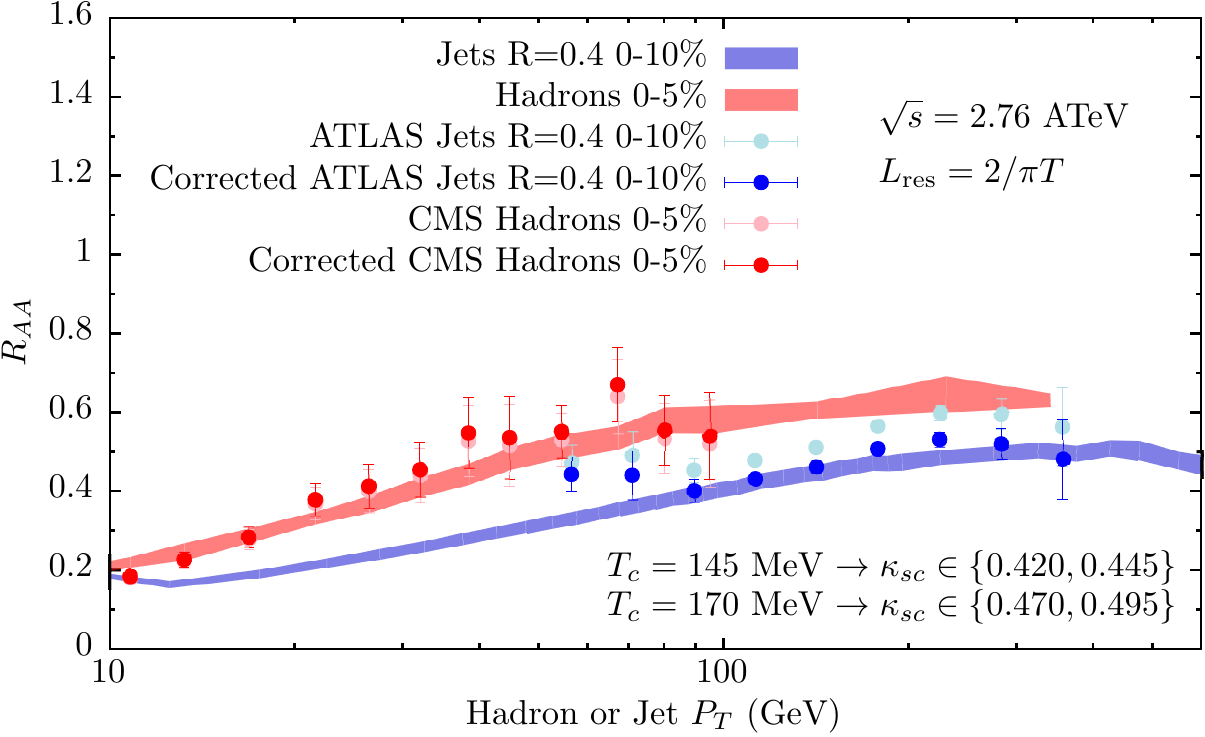}
\caption{\label{fig:visual} Results for $\hraa$ and $\jraa$ from our model with its parameter fixed via the global fit, compared to CMS \cite{CMS:2012aa} and ATLAS \cite{Aad:2014bxa} data. Error bars on the experimental data points show only the uncorrelated error.
The  corrected data points have been shifted according to the best fit value of the correlated error correction ~\cite{Adare:2008qa}.
 Colored bands show results from the hybrid model with $\lres=2/(\pi T)$, with the bands spanning results obtained with $T_c=145$ and 170 MeV, in each case using the value of $\aSC$ obtained from the global fit in Fig.~\ref{fig:fits}}
\end{figure}

In Fig.~\ref{fig:visual} we provide an impression of how individual points in Fig.~\ref{fig:fits} are obtained by showing
a subset of our results compared to data for  $\jraa$ with anti-$k_t$ radius of $R=0.4$~\cite{Cacciari:2008gp}, and $\hraa$ (plotted together, meaning that the horizontal 
axis corresponds to either hadron or jet $p_T$) for PbPb collisions with $\sqrt{s_{\rm NN}}=2.76$~TeV at the LHC. The bands from the model comprise the results obtained for the $2 \sigma$ range for $\aSC$ as extracted from the global fits for both values of $T_c$, and using $\lres=2/(\pi T)$.

\begin{figure}[t]
\includegraphics[scale=0.72]{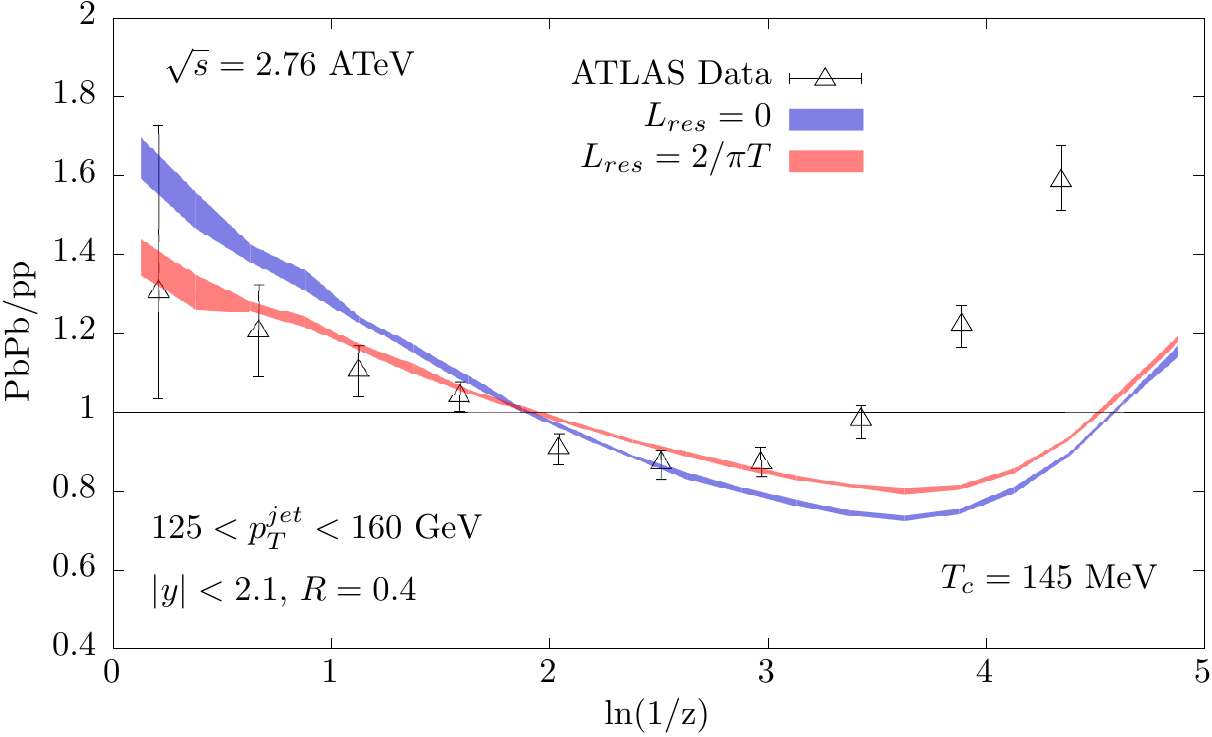}
\caption{\label{fig:allFF} The ratio of fragmentation functions for jets in PbPb collisions with 
$125$~GeV~$ < p_T^{\rm jet} < 160$~GeV to those for jets with the same $p_T$  in pp collisions, with hybrid model predictions for $\lres=0$ and $2/(\pi T)$ 
compared to ATLAS data~\cite{Aaboud:2017bzv}. 
The enhancement in the PbPb fragmentation functions for $z$ near 1, on the left, shows that
jets in PbPb collisions are more likely to feature a few hard fragments that each carry a significant fraction of the whole jet momentum than is the case for jets in pp collisions.
Note that, in this observable, we do see some evidence favoring $\lres=2/{\pi T}$ over
$\lres=0$.
The disagreement between the hybrid model predictions and data at small $z$, on the right, points to the need to improve the current hybrid model implementation~\cite{Casalderrey-Solana:2016jvj} of the wakes that jets deposit in the medium.}
\end{figure}
\begin{figure}[t]
\includegraphics[scale=0.72]{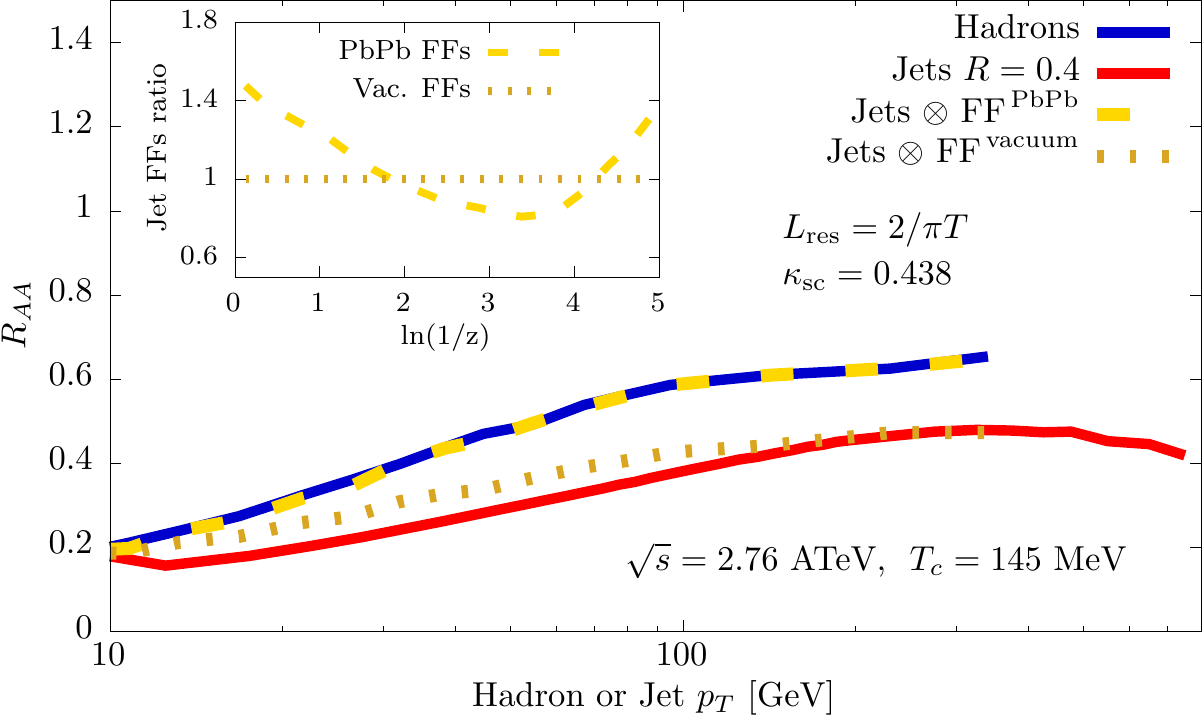}
\caption{\label{fig:convwithFF} Hybrid model results for $\hraa$ (solid blue) and $\jraa$ (solid red) in LHC collisions with $\sqrt{s_{\rm NN}}=2.76$~TeV.
Hadron spectra can be obtained from jet spectra by convolving the jet spectra
with a jet fragmentation function. Here, we do so in two ways.
First, by consistently using the PbPb and pp fragmentation functions from the hybrid model calculations themselves (dashed yellow in the inset), we do indeed recover the hybrid model result for $R_{\rm AA}^{\rm hadron}$  
from the hybrid model result for  $R_{\rm AA}^{\rm jet}$. If instead, in dotted yellow,
we wrongly assume that the quenched fragmentation functions are unmodified from the
vacuum ones, we obtain an incorrect, but interesting, result.  
}
\end{figure}

\textit{Modification of jet fragmentation functions.} 
Following the discussion in the Introduction, we turn now to jet fragmentation functions.
By definition, fragmentation functions 
count the mean number of hadrons, per jet, that carry a fraction $z$ of the whole jet energy, 
with $z$ usually defined in experimental analyses as $z\equiv (\bold{p}_{\rm h} \cdot \bold{p}_{\rm j})/|\bold{p}_{\rm j}|^2$, where $\bold{p}_{\rm h}$ 
and $\bold{p}_{\rm j}$ are the three-momentum of the hadron and jet, respectively.
The ratio of fragmentation functions in PbPb and pp collisions
was introduced as an observable that is affected by jet quenching 
in Ref.~\cite{Chatrchyan:2014ava} 
and has been measured by both CMS and ATLAS~\cite{Chatrchyan:2014ava,Aaboud:2017bzv,Aaboud:2018hpb}.
Here, we are interested in the enhancement in this ratio close to $z\sim 1$~\footnote{The enhancement in this ratio for soft particles with 
$p_T \lesssim 3$ GeV and hence small $z$ is of interest for (at least) two reasons that are
unrelated to our considerations. 
This enhancement may receive a contribution from 
loss of color coherence induced by multiple scatterings with the medium \cite{Mehtar-Tani:2014yea,Caucal:2018dla}. 
And, assuming that the wake left behind in the plasma by the jet becomes soft
hadrons in this momentum range (which is likely) which carry the momentum lost
by the jet (which is necessary by momentum conservation) then the wake must translate
into an enhancement in the fragmentation function ratio in this soft regime~\cite{Casalderrey-Solana:2016jvj,Chen:2017zte}. 
}.
As we described in the Introduction, due to the steeply falling jet spectrum 
whenever we trigger on a high $p_T$ hadron we are biasing our sample towards narrow jets that fragmented into  
few, hard, hadrons.
We see from the fragmentation function ratio near $z\sim 1$ in 
Fig.~\ref{fig:allFF} that such jets are more common in PbPb collisions than in pp collisions.
While the first results from ATLAS at $\sqrt{s_{\rm NN}}=2.76$ TeV already showed hints of an enhancement in this ratio at high $z$~\cite{Aaboud:2017bzv}, this 
behavior has been convincingly demonstrated by the recent precise measurements reported in Ref.~\cite{Aaboud:2018hpb} for $\sqrt{s_{\rm NN}}=5.02$ TeV. 
The agreement between our hybrid model calculations, which predate Ref.~\cite{Aaboud:2018hpb}, and these measured data suggests
that this enhancement in the probability for finding hard fragmenting jets 
has the same origin as the lesser suppression of hadron yields relative to jet yields that our 
model also describes.
We shall confirm this quantitatively in Fig.~\ref{fig:convwithFF}.

In Fig.~\ref{fig:convwithFF},
we show hybrid model calculations 
of $\hraa$  and $\jraa$ for anti-$k_t$ radius $R=0.4$, in 
collisions with $\sqrt{s_{\rm NN}}=2.76$ TeV
with $\aSC$ set to its best fit value for $T_c=145$~MeV and $\lres=2/(\pi T)$, namely $\aSC=0.438$.
By convolving the PbPb (pp) jet spectrum with the appropriately binned fragmentation functions obtained in PbPb (pp) collisions whose ratio is depicted by the dashed yellow curve in the inset, one can recover the corresponding hadronic spectra and, in particular, the ratio of medium over vacuum spectra, as can be seen via the agreement between the dashed yellow curve in the main panel of Fig.~\ref{fig:convwithFF} and the solid blue one. This had to work out, since the dashed yellow, red and blue curves are taken from the same hybrid model calculation.
%
%
The most interesting comparison in Fig.~\ref{fig:convwithFF} comes when 
we (incorrectly) assume that the jet fragmentation function in PbPb collisions is the same as in pp collisions, as in the dotted yellow curve in the inset.
We see that upon making this assumption we completely lose the ability to explain the difference between hadron and jet suppression, with the dotted yellow curve in the main panel showing that when the jet spectrum is convolved with this (incorrect) PbPb fragmentation function, the resulting (incorrect) hadron 
spectrum is rather similar to the jet spectrum.
%
What we learn from this is that the 
difference between the suppression of hadron yields and jet yields, with $\hraa > \jraa$ seen in experiments and
in the hybrid model, is equivalent to the presence of a high-$z$ enhancement in the fragmentation function ratio.

\textit{Conclusions.}  The enhancement in the ratio of fragmentation functions in PbPb and pp collisions at high-$z$ was  predicted using the hybrid strong/weak coupling model~\cite{Casalderrey-Solana:2016jvj}.
It originates from the fact that wider jets containing more partons 
at large angles on average lose more
energy than narrower jets. 
If two jets have the same momentum, the one that is narrower, meaning that it has 
a smaller jet mass, 
will lose less energy than the wider one made up of more fragments that are sufficiently separated from each other that they each lose energy independently.
%
This effect, together with the steeply falling jet spectrum, means that
selecting a sample of jets with a given energy in PbPb collisions results in a  bias toward finding 
narrow, hard fragmenting, jets. 
This mechanism thus also leads to
 the  enhancement of the fragmentation
function at high-$z$, as measured in experiments.

The same effect also means that since when we select a sample of high-$p_T$ hadrons
we are selecting hadrons that come from unusually narrow jets with unusually hard fragmentation, we are selecting hadrons from jets that lose less energy than typical jets do.  Hence, hadron yields are less suppressed in PbPb collisions than jet yields are.

In support of these conclusions, we have seen that at the  same time that the hybrid model provides a good description
of the fragmentation function ratio at high-$z$, it provides a simultaneous description
of hadron and jet suppression in heavy ion collisions.



\textit{Acknowledgments.}~This work was supported by  grants SGR-2017-754, FPA2016-76005-C2-1-P and MDM-2014-0367, by Funda\c c\~ao para a Ci\^encia e a Tecnologia (Portugal) contracts CERN/FIS-PAR/0022/2017 and IF/00563/2012, by US NSF grant ACI-1550300 and by US DOE Office of Nuclear Physics contract DE-SC0011090.  JCS is a Royal Society University Research Fellow (on leave). KR and GM gratefully acknowledge the hospitality of the CERN theory group. We thank
Carlota Andr\'es, Peter Jacobs, Yen-Jie Lee, Gunther Roland,  Carlos Salgado, Wilke van der Schee, Marco van Leeuwen, Xin-Nian Wang, Urs Wiedemann and Korinna Zapp for helpful conversations.



\end{document}